\documentclass[aps,amsfonts,eqsecnum,superscriptaddress,nofootinbib,longbibliography,notitlepage]{revtex4-1}

\usepackage{graphicx}
\usepackage{xcolor}
\usepackage{rotating}
\usepackage{float}
\usepackage{caption}
\usepackage{subcaption}
\usepackage{amsmath,amssymb,graphics,amsthm,isomath}
\usepackage[colorlinks=true, urlcolor=violet, linkcolor=blue, citecolor=red, hyperindex=true, linktocpage=true]{hyperref}
\usepackage[capitalise,compress]{cleveref}

\makeatletter
\renewcommand{\p@subsection}{}
\renewcommand{\p@subsubsection}{}
\makeatother

\usepackage{xcolor}
\usepackage{mathtools}
\usepackage[section]{placeins}

\usepackage{dsfont}

\begin{document}

\title{Holographic subdiffusion}

\author{Koushik Ganesan}
\email{koushik.ganesan@colorado.edu}
\affiliation{Department of Physics and Center for Theory of Quantum Matter, University of Colorado, Boulder CO 80309, USA}

\author{Andrew Lucas}
\email{andrew.j.lucas@colorado.edu}
\affiliation{Department of Physics and Center for Theory of Quantum Matter, University of Colorado, Boulder CO 80309, USA}

\begin{abstract}
We initiate a study of finite temperature transport in gapless and strongly coupled quantum theories with charge and dipole conservation using gauge-gravity duality. In a model with non-dynamical gravity, the bulk fields of our model include a suitable mixed-rank tensor which encodes the boundary multipole symmetry.  We describe how such a theory can arise at low energies in a theory with a covariant bulk action. Studying response functions at zero density, we find that charge relaxes via a fourth-order subdiffusion equation, consistent with a recently-developed field-theoretic framework.
\end{abstract}

\date{\today}

\maketitle
\tableofcontents

\section{Introduction}
The past decade has seen enormous work on the ``AdS/CMT" correspondence, whereby the  condensed matter physics of certain strongly coupled quantum systems can be studied through a holographically dual classical gravity theory:  see \cite{Hartnoll:2009sz,McGreevy:2009xe,zaanenbook,hartnoll2016holographic} for reviews.  Much of this work relies on a ``bottom-up" approach, whereby one simply posits a bulk action which contains the appropriate symmetry and field content of the desired strongly coupled theory, without knowing the precise field theory dual.  The goal of this approach is not to make quantitative predictions about any particular condensed matter model, but rather to guide our intuition for how to think about and organize our understanding of strongly coupled systems without quasiparticles.

In this paper, we initiate a holographic study of models of strongly coupled constrained quantum dynamical systems.  Such constrained dynamics first arose in the study of models with microscopic excitations called fractons (which are individually immobile, but collectively mobile) \cite{chamon2005quantum, haah2011local, vijay2015new, prem2017emergent, prem2018cage, slagle2017fracton, slagle2017quantum, pretko2017emergent, pretko2017higher, you2018subsystem, you2018symmetric, schmitz2018recoverable, moudgalya, sous2019fractons}; see also the reviews \cite{nandkishore2018fractons,pretko2020fracton}.  However, such constrained dynamics can arise in a variety of other experimentally-relevant contexts \cite{Guardado_Sanchez_2020}.  A significant challenge with studying the dynamics of a many-body system with constraints is that, essentially by definition, there is not an obvious quasiparticle description -- after all, if individual excitations cannot move in space, the only possible way for dynamics to proceed is through the interactions of multiple excitations.  Many of the systems which have been amenable to study in the past are models of random unitary circuits (with constraints), yet typically the models studied amount to classical Markov chains \cite{pai2019localization, khemani2019local, 2020PollmannFragmentation, morningstar2020kineticallyconstrained,feldmeier2020anomalous}, which should be sufficiently generic to capture hydrodynamic phenomena, but may not capture quantum transport phenomena and the crossover away from hydrodynamics at shorter time and length scales.

Here, we attack this problem via holography.  Our first main result is a prescription for how to model such dynamics using holography, including from a covariant description in the bulk.  Our construction follows from the realization that fracton matter (and constrained dynamics more generally) are described by theories invariant under a multipole algebra \cite{gromov2018towards}, and naturally couple to higher-rank gauge fields \cite{xu2006novel, xu2010emergent, rasmussen2016stable, pretko2017generalized, pretko2017subdimensional, Seiberg:2020bhn,Seiberg:2020cxy,Seiberg:2020wsg}.

We then proceed to analyze the linear response of such theories.  In systems which are charge neutral, we focus on the correlation functions of the conserved density and the (higher rank) current operators.  In the long wavelength limit, our results are in complete agreement with the universal framework of \cite{gromov2020fracton}.  We find that some aspects of more conventional holographic transport, such as the membrane paradigm \cite{1986bhmp.book.....T,Iqbal_2009}, extend to our theory, while others, such as the relationship between chaos and diffusion \cite{Blake_2016, Blake:2016sud,Lucas:2016yfl,Blake:2016jnn,Blake:2017qgd,Kim:2017dgz,Baggioli:2017ojd,Ahn:2017kvc}, do not apply to the subdiffusion constant.  Holography also allows us to calculate response functions at short wavelengths, where hydrodynamics breaks down.  We numerically calculate transport coefficients across a ``hydrodynamic-to-collisionless" crossover, where the high frequency physics is effectively captured by a zero temperature scale-invariant limit, whose properties are easily computed.  Of course, here the ``collisionless" regime of a conventional kinetic theory is replaced by a scale invariant field theoretic limit, where the scaling of the conductivity is (mostly) fixed by dimensional analysis.


\section{General field theoretic considerations}
\subsection{Background gauge fields and the Ward identity}
We begin by thinking, from a general field theoretic perspective, about how to build actions which encode a dipole conservation law.  Since this is an unusual problem, let us begin by reviewing a simpler problem, with an ordinary (unbroken) U(1) symmetry (i.e. charge conservation but not dipole conservation).  In this case, we are free to \emph{locally} rotate the U(1) phase $\phi(x)$ if we are willing to introduce a classical background gauge field $A_\mu$ to absorb the rotation, via $A_\mu \rightarrow A_\mu + \partial_\mu \phi$.  The invariance of the action under this local rotation then implies that 
\begin{equation}
  0=  \frac{\delta Z}{\delta \phi(x)} = \partial_\mu \frac{\delta Z}{\delta A_\mu(x)}  = \partial_\mu J^\mu
\end{equation}
where $S$ denotes the generating functional of the field theory and $J^\mu$ denotes a local current operator.  Clearly, this is the local Ward identity corresponding to charge conservation.   As usual, $\mu\nu\cdots$ indices refer to spacetime coordinates; in a holographic context, they do not include the extra bulk radial dimension.

How should we generalize this to account for local dipole conservation? A heuristic argument is as follows:  there is a non-trivial choice of phase $\phi$ in the charge conserving theory: \begin{equation}
    \phi = \mu t
\end{equation}
which allows us to study the theory above at a chemical potential $\mu$, since (despite being pure phase) this $\phi$ leads to \begin{equation}
    A_t = \mu.
\end{equation}
In a dipole conserving theory, we would like to be able to impose a chemical potential $\tilde \mu_i$ for the dipole moment as well:
\begin{equation}
A_t = \mu + \tilde \mu_i x_i    .  \label{eq:Atdipole}
\end{equation}
Here $i,j\cdots$ represent spatital indices in each of $d$ spatial directions.  However, since $\partial_i A_t - \partial_t A_i \ne 0$ for the ansatz (\ref{eq:Atdipole}), we cannot impose such a dipole chemical potential if we couple the theory to the usual gauge fields $A_\mu$.  In fact, the kind of structure which we need to couple to is a ``mixed rank" gauge field \cite{pretko2017generalized, pretko2017subdimensional} \begin{subequations} \label{eq:localdipoletrans}\begin{align}
    A_t &\rightarrow A_t + \partial_t \phi, \\
    A_{ij} &\rightarrow A_{ij} - \partial_i\partial_j\phi.
\end{align}\end{subequations}
Note that $A_{ij}=A_{ji}$ is symmetric.  Since the gauge invariant object here is $\partial_i\partial_j A_t + \partial_t A_{ij}$, we can now see that (\ref{eq:Atdipole}) is acceptable.  

Hence, we must couple the dipole-conserving field theory to a mixed rank gauge field: \begin{equation}
    Z[A_t,A_{ij}] = \left\langle \exp\left[\mathrm{i}\int \mathrm{d}^dx\mathrm{d}t \; \left(A_t J^t + A_{ij}J^{ij}\right) \right]  \right\rangle .
\end{equation}
Moreover, since $A_t$ and $A_{ij}$ are the fields which couple to observables, we conclude that the appropriate observables of this theory are $J^t$ and $J^{ij}$.  Moreover, the invariance of the theory under the local transformation (\ref{eq:localdipoletrans}) leads to \begin{equation} \label{eq:ward}
    0 = \frac{\delta Z}{\delta \phi(x)} = \partial_t J^t + \partial_i\partial_j J^{ij} = 0.
\end{equation}

\subsection{Subdiffusion}\label{sec:subdifffield}
It is natural to ask, given the unusual form of charge conservation in (\ref{eq:ward}), what is the emergent hydrodynamics of a theory with dipole conservation.  Hydrodynamics is an effective theory describing long time (at times longer than a ``collision mean free time", or more abstractly ``thermalization time") behavior of many-body dynamics with conserved quantities \cite{Haehl:2015foa,Crossley:2015evo,Jensen:2017kzi}; it can also be developed for systems with almost conserved quantities \cite{Grozdanov_2019}.   It is a powerful framework because it is sensitive only to the symmetries of the system, and hence a large class of microscopically diverse systems fall into the same hydrodynamic universality class.

The hydrodynamics of systems with multipole conservation laws was systematically developed in \cite{gromov2020fracton}, assuming that the charge dynamics decouples from other degrees of freedom.  It was shown that in these classes of systems that the dynamics was sub-diffusive with unique sub-diffusive exponents. In particular, with dipole conservation, one finds that \begin{equation}
    J_{ij} = B_1 \partial_i \partial_j \rho + B_2 \delta_{ij} \partial_k\partial_k \rho + \cdots
\end{equation}
where $\cdots$ denotes higher derivative corrections.  Hence, the resulting equation of motion reads \begin{equation}
    \partial_t \rho = -(B_1+B_2)(\partial_i\partial_i)^2 \rho.
\end{equation}
In other words, there will be a quasinormal mode (a pole in Green's functions) at frequency $\omega = -\mathrm{i}(B_1+B_2)k^4$. 

Observe that the electric field in the dipole-conserving theory is defined by \begin{equation}
    E_{ij} = -\partial_t A_{ij} - \partial_i\partial_j A_t,
\end{equation}
and so the conductivity we are looking for is a fourth rank tensor: \begin{equation}
    \langle J_{ij}\rangle = \sigma_{ij,kl}E_{kl}.
\end{equation}
Here we are sloppy about raising and lowering indices, as we are focused on field theories on flat space.

\section{Holographic model}

\subsection{A bottom-up bulk model}
With this most basic introduction to the problem of dipole-conserving hydrodynamics, we are now ready to design a holographic bulk theory capable of encoding this physics.  Following the usual holographic dictionary, the dynamical bulk fields will correspond to the source fields which couple to operators of interest in the field theory: these will be the energy-momentum tensor $T^{\mu\nu}$, the charge density $J^t$ and the dipole current $J^{ij}$.  These fields couple to the metric $g_{ab}$, and (ideally) the mixed-rank gauge field $A_t$ and $A_{ij}$, respectively.  Note that $ab\cdots$ indices are used to refer to \emph{all} dimensions: $d$ boundary spatial dimensions, time, and the bulk radial dimension in holography.

There are two important subtleties that we must immediately address.  Above, we listed the metric $g_{ab}$ as a bulk field dual to energy/momentum:  this field carries bulk spatial indices.  Indeed, all of the non-trivial structures of holography rely on equations of motion that depend on bulk coordinates.  Yet at the same time, only the boundary spacetime components of $g_{\mu\nu}$ carry information about the couplings in the boundary theory (and can therefore be used to read off response functions).  These observations are reconciled by noting that the bulk actions are invariant under diffeomorphisms, which allow us to ``gauge fix" $g_{ar}$  (here $r$ represents the bulk radial coordinate).  This is very well understood \cite{Hartnoll:2009sz,McGreevy:2009xe,zaanenbook,hartnoll2016holographic}.  However, we are now trying to build a holographic theory with a mixed-rank tensor whose boundary indices are $(A_t,A_{ij})$. How is one to make sense of a mixed rank tensor in a geometric theory of gravity, where the metric can mix space and time indices?  And how do we handle the extra dimension? 

These issues are not merely ``mathematical", but arise from a physical issue.  When trying to put a dipole-conserving theory on curved space, the multipolar conservation law can be destroyed due to spatial curvature \cite{gromovprl,Slagle:2018kqf,doshi2020vortices}: dipoles rotate when parallel transported around loops!  From our perspective, this is not really an issue -- we are not interested (at least in this paper) in putting our \emph{field theory} on a curved spacetime; only the bulk is curved.  However, since this issue spoils existing attempts to couple a mixed rank gauge theory to curved space, there is no ``Maxwell action" for mixed-rank tensors that we can write down in the bulk.


Hence we proceed first along phenomenological grounds.  First, we note that the UV theory (at least from the point of view of the multipole-conserving dynamics) is clearly non-relativistic, since the operators $(J_t,J_{ij})$ certainly do not form a covariant first or second rank tensor. Hence the effective holographic theory capturing the multipole-conserving dynamics will not appear covariant, at least in a relativistic theory of gravity.  Generating a non-relativistic geometry has been extensively studied \cite{Son:2008ye,Balasubramanian:2008dm,Taylor:2008tg,Goldstein:2009cv,Charmousis:2010zz,Dong:2012se} in the holographic literature.  In each approach, there must be background fields (such as Weyl tensors or background gauge fields) which pick out the $r$ and $t$ directions as distinct from the spatial directions $i$. 
So we can, in principle, use these objects to pick out the $r$ and $t$ directions as distinct, and our effective action need not be written in terms of manifestly covariant objects, so long as the background metric is not dynamical.

While a non-relativistic theory of gravity \cite{Janiszewski:2012nb,horava} may be a better starting point, as a matter of convenience, the theory of relativistic gravity is more broadly understood, so we would like to ask whether it might be possible to interpret \eqref{3} in some relativistic theory of gravity, albeit in a background which explicitly breaks Lorentz symmetry. We will return to this issue in the next subsection.

In a non-dynamical geometry of the kind described in Section \ref{sec:dictionary}, we desire the remaining gauge fields to be invariant under the combined gauge transformation
\begin{subequations}\label{1}
\begin{align}
    A_{ij}&\xrightarrow[]{}A_{ij}-\partial_i\partial_j\lambda \\
    A_{t}&\xrightarrow[]{}A_{t}+\partial_t\lambda\\
    A_r&\xrightarrow[]{}A_r+    \partial_r\lambda
\end{align}
\end{subequations}
There are not covariant derivatives in this expression because, as noted above, the field theory does not live on curved space.  Even in the bulk, we propose that (\ref{1}) holds, and will see how this can (approximately) arise in a non-dynamical geometry in the following subsection. We can then build the following gauge-invariant objects: 
\begin{subequations}\label{2}
\begin{align}
    F^1 &= \partial_tA_r-\partial_rA_t\\
    F^2_{ij} &= \partial_tA_{ij} + \partial_i\partial_jA_t\\
    F^3_{ij} &= \partial_rA_{ij} + \partial_i\partial_jA_r\\
    F^4_{ijk} &= 2\partial_kA_{ij}-\partial_iA_{jk}-\partial_jA_{ki}
\end{align}
\end{subequations}
We are interested in theories which are charge conjugation symmetric, as well as working within a linear response regime, so it is sufficient to include only quadratic terms in the following phenomenological holographic bulk action:
\begin{equation}\label{3}
\begin{split}
    S =\frac{1}{2} \int d^{d+2}x\sqrt{-g}\bigg(C_0g^{rr}g^{tt}(F^1)^2+C_1g^{tt}g^{jj}g^{ii}(F^2_{ij})^2+C_4g^{rr}g^{jj}g^{ii}(F^3_{ij})^2+C_2g^{tt}g^{ii}g^{jj}F^2_{ii}F^2_{jj}+\\C_5g^{rr}g^{ii}g^{jj}F^3_{ii}F^3_{jj}+C_3g^{kk}g^{jj}g^{ii}(F^4_{ijk})^2\bigg)
\end{split}
\end{equation}
where $C_0,C_1,C_2,C_3,C_4,C_5$ are dimensionless constants.  We do not allow these coefficients to depend on a bulk dilaton field (and thus effectively on $r$), although that is a natural generalization of our work.  We can set $C_0 = 1$ by rescaling the field $A$.  We also set $C_4 = C_1$ and $C_5 = C_2$.  If this choice is not made, then infalling plane waves in Kruskal coordinates are not solutions to the near-horizon equations of motion in the presence of a black hole.   Lastly, in the rest of the paper, we work in the gauge $A_r=0$.  


\subsection{Covariant bulk action}\label{lkj}
In this subsection, we elaborate on how one can recover a model equivalent to (\ref{3}) from an explicitly covariant bulk action, coupled to Einstein gravity.  The equivalence will hold within the regime of linear response, and without dynamical gravity; with dynamical gravity, the equivalence can explicitly break and may signal subtleties about fracton hydrodynamics that are not understood on field theoretic grounds. 

We assume the presence of a background dilaton scalar field $\Phi$, which partially supports the background hyperscaling violating metric.  Then we introduce $d$ scalar fields $\phi^I$, sometimes referred to as ``axions" due to an assumed shift symmetry $\phi^I \rightarrow \phi^I + c$.  We will choose equations of motions for $\phi^I$ so that a consistent solution of the bulk equations of motion is 
\begin{equation}
    \phi^I = x^i\delta^I_i
\end{equation}
where the index $I\in \{1,2,..d\}$ is a field index, and does not transform under diffeomorphisms.  The \emph{boundary conditions}, not the bulk action, mix the $I$ and $i$ indices.  This construction has also arisen in the effective theory of ideal fluids \cite{Dubovsky:2011sj} and in holographic models of momentum relaxation \cite{Andrade:2013gsa}.

We also put in $d+1$ U(1) vector gauge fields denoted by $A_\mu$ and $A^I_\mu$. Now using our scalar fields, we can construct projectors $P_{ab}$ and $Q_{ab}$ given by
\begin{subequations}
\begin{align}
    Q_{ab} &= \nabla_a\phi^I\nabla_b\phi_IG(\Phi)\\
    P^a_b &= g^a_b - Q^a_b
\end{align}
\end{subequations}
where the indices $a$ and $b$ are raised and lowered with the metric tensor $g$.  We choose the function \begin{equation}
    G(\Phi(r)) = g_{xx}(r),
\end{equation}
which can be upheld so long as the metric is non-dynamical, at least in our bottom up model.  The  projector $P$ is used to project onto $r$ and $t$ components as $P^a_b \neq 0$ only when $a=b\in\{r,t\}$. Similarly $Q^a_b$ projects onto the the spatial components.
Now we propose the following Lagrangian:
\begin{align}\label{lk}
    L \supset P^b_dF_{bc}P^c_eF^{de}+U(\Phi)\nabla_a\phi^I\nabla^a\phi_JF^I_{bc}F^{J,dc}Q^b_d+V(\Phi)\nabla_a\phi^IF^{ac}_IP^d_c\nabla^b\phi_JF^J_{bd}+Y(\Phi)Q_{ab}A^aA^b\nonumber\\+W(\Phi)(\nabla^c\phi_JA_{I,c}-\nabla^c\phi_IA_{J,c})(\nabla_d\phi^JA^{I,d}-\nabla_d\phi^IA^{J,d})+Z(\Phi)M^I_bM^a_IP^b_a
\end{align}
where $F_{ab} = \nabla_a A_b - \nabla_b A_a$, $F^I_{ab} = \nabla_a A^I_b - \nabla_b A^I_a$, and \begin{equation}
    M^I_b = \nabla_a\Phi^I\nabla^aA_b-\nabla_a\phi^I\nabla_bA^a+A^J_b\nabla_a\phi_J\nabla^a\phi^I.
\end{equation} The terms proportional to $Y$, $W$, and $Z$ are used to enforce the following constraints respectively --
\begin{subequations}\label{bk}
\begin{align}
    A_i &\approx 0\\
   \delta^I_i A_{I,j} &\approx \delta^J_j A_{J,i}\\
   \delta^I_i A_{I,a} &\approx -\partial_i A_{a}\;\;\;\; (a\in\lbrace r,t\rbrace)
\end{align}
\end{subequations}
respectively with $i$, $j$ running over spatial coordinates.  These constraints can be understood from the equations of motion for the fields $A_i$, $A_{I,j}$, $A_{I,t}$ and $A_{I,r}$ respectively:
\begin{subequations}
\begin{align}
     Y(\Phi)g^{ii}A_i+\partial_r(Z(\Phi)g^{rr}(\partial_iA_r-\partial_rA_i+A_{i,r}))+\partial_t(Z(\Phi)g^{tt}(\partial_iA_t-\partial_tA_i+A_{i,t})) &= 0\\
     2W(\Phi)(A_{i,j}-A_{j,i})-\partial_t(U(\Phi)g^{ii}g^{jj}g^{tt}(\partial_tA_{i,j}-\partial_jA_{i,t}))-\partial_r(U(\Phi)g^{ii}g^{jj}g^{rr}(\partial_rA_{i,j}-\partial_jA_{i,r})&)\nonumber\\-\delta_{ij}(\partial_t(V(\Phi)g^{ii}g^{jj}g^{tt}(\partial_tA_{i,i}-\partial_iA_{i,t}))+\partial_r(V(\Phi)g^{ii}g^{jj}g^{rr}(\partial_rA_{i,i}-\partial_iA_{i,r}))) &= 0\\
     Z(\Phi)g^{tt}(\partial_iA_t-\partial_tA_i+A_{i,t})+\partial_j(U(\Phi)g^{ii}g^{jj}g^{tt}(\partial_tA_{i,j}-\partial_jA_{i,t}))+\partial_i(V(\Phi)g^{ii}g^{jj}g^{tt}(\partial_tA_{jj}-\partial_jA_{j,t})) &= 0\\
     Z(\Phi)g^{rr}(\partial_iA_r-\partial_rA_i+A_{i,r})+\partial_j(U(\Phi)g^{ii}g^{jj}g^{rr}(\partial_rA_{i,j}-\partial_jA_{i,r}))+\partial_i(V(\Phi)g^{ii}g^{jj}g^{rr}(\partial_rA_{jj}-\partial_jA_{j,r})) &= 0
\end{align}
\end{subequations}
where $i$, $j$ run over spatial indices. Now suppose that $Y$, $W$ and $Z$ are sufficiently large.  Then observe that the first terms in each of the equations above serve to approximately enforce the constraints in (\ref{bk}).   Alternatively, when $Y$, $Z$ and $W$ are large, the action will oscillate too rapidly unless the arguments vanish.
We conclude that the first three terms of \eqref{lk} remain non-trivial, and are given by 
\begin{subequations}
\begin{align}
    P^b_dF_{bc}P^c_eF^{de} &= 2g^{tt}g^{rr}F^2_{tr}\\
    U(\Phi)\nabla_a\phi^I\nabla^a\phi_JF_{I,bc}F^{J,dc}Q^b_d &= U(\Phi)\sum^d_{I,J=1}g^{II}g^{JJ}(g^{tt}F^2_{I,tJ}+g^{rr}F^2_{I,rJ})\approx U(\Phi)\sum^d_{I,J=1}g^{II}g^{JJ}(g^{tt}(\partial_tA_{I,J}+\partial_I\partial_JA_{t})^2\nonumber\\&\hspace{20em}+g^{rr}(\partial_rA_{I,J}+\partial_I\partial_JA_r)^2) \\
    V(\Phi)\nabla_a\phi^IF^{ac}_IP^d_c\nabla^b\phi_JF^J_{bd} &= V(\Phi)\sum_{a=\{r,t\}}g^{aa}\bigg(\sum^d_{I=1}g^{II}F_{I,aI}\bigg)\bigg(\sum^d_{J=1}g^{JJ}F_{J,aJ}\bigg)\approx V(\Phi)\sum^d_{I,J=1}g^{II}g^{JJ}\bigg(g^{tt}(\partial_tA_{I,I}\nonumber\\&\hspace{2em}+\partial_I\partial_IA_{t})(\partial_tA_{JJ}+\partial_J\partial_JA_t)+g^{rr}(\partial_rA_{I,I}+\partial_I\partial_IA_r)(\partial_rA_{J,J}+\partial_J\partial_JA_r)\bigg) 
\end{align}
\end{subequations}
where we have taken into account the constraint given in \eqref{bk}. We see that these are indeed terms of our original action \eqref{3} provided we treat the metric to be non-dynamical. 

At finite density the metric will couple to the gauge field fluctuations.  Under this construction an extra term given by $g^{tx}g^{rr}F_{tr}F_{xr}$ arises at linear order.  This term appears to be demanded holographically.  A field theoretic interpretation of this term seems to be that at finite density, there must be a contribution to the charge current $J^x$, since the cross-susceptibility $\chi_{J_xP_x} \ne 0$ \cite{hartnoll2016holographic}.  It would be interesting to better understand fracton hydrodynamics at finite density, whether through holography or general field theoretic considerations.

\subsection{Holographic dictionary} \label{sec:dictionary}

In order to describe the holographic dictionary, we must now describe the UV behavior of the background geometry.  Let us assume it takes the hyperscaling-violating Lifshitz form \cite{Dong:2012se,Gouteraux:2011ce}
\begin{equation}\label{789}
    \mathrm{d}s^2 = L^2\bigg(\frac{r}{R}\bigg)^{\frac{2\theta}{d}}\bigg(-\frac{\mathrm{d}t^2}{r^{2z}}+\frac{\mathrm{d}r^2}{r^2}+\frac{\mathrm{d}\Vec{x}^2}{r^2}\bigg)
\end{equation}
where $z$ is known as the dynamical critical exponent, and $L$ is the AdS radius. Unlike AdS metric, these geometries treat time and spatial components on different footing by generalizing the scaling symmetry $\{t,\Vec{x}\}\xrightarrow[]{}\{\lambda^zt,\lambda\Vec{x}\}$. They preserve translational, time-reversal and rotational symmetry but break boosts.  When $z=1$, we restore the full isometry group of AdS: $\mathrm{SO}(d+1,2)$. Although these geometries are not solutions to pure Einstein's gravity, they arise as solutions to theories such as Einstein-Maxwell-dilaton theory \cite{Charmousis:2010zz} and higher derivative gravity \cite{Ay_n_Beato_2009}.
 For these metrics we assume $0\leq\theta\leq d-1$ and $z\geq1+\frac{\theta}{d}$ \cite{Huijse:2011ef}, which can be shown to satisfy the null energy condition, as well as exhibit relatively conventional ground state entanglement.  If we are to work at finite temperature then we need to add an emblackening factor to the metric as follows
 \begin{equation}\label{789}
    \mathrm{d}s^2 = L^2\bigg(\frac{r}{R}\bigg)^{\frac{2\theta}{d}}\bigg(-f(r)\frac{\mathrm{d}t^2}{r^{2z}}+\frac{\mathrm{d}r^2}{f(r)r^2}+\frac{\mathrm{d}\Vec{x}^2}{r^2}\bigg)
\end{equation}
where $f(r)\xrightarrow[]{}1$ as $r\xrightarrow[]{}0$ and $f(r_+)=0$ corresponds to the horizon. The functional form of $f$ is \begin{equation}
    f(r) = 1-\bigg(\frac{r}{r_+}\bigg)^{d+z-\theta},
\end{equation}
and this can be found by carefully solving the bulk equations of motion in a suitable theory of dynamical gravity: see \cite{hartnoll2016holographic}.

Now, let us describe the physics of the non-covariant holographic model (\ref{3}). Assuming a static background metric, the bulk equations of motion for the components $A_r$, $A_t$, and $A_{ij}$ of the gauge field are given by 
\begin{subequations}\label{1001}
\begin{align}
    0&=g^{rr}g^{tt}\partial_t(\partial_rA_t)+\sum_{ij}C_1g^{rr}g^{ii}g^{jj}\partial_i\partial_j\partial_rA_{ij}+\sum_{ij}C_2g^{rr}g^{ii}g^{jj}\partial_j\partial_j\partial_rA_{ii}\\
    0&=-\partial_r(\sqrt{-g}g^{rr}g^{tt}\partial_rA_t)+\sum_{ij}C_1\sqrt{-g}g^{tt}g^{ii}g^{jj}\partial_i\partial_j(\partial_tA_{ij}+\partial_i\partial_jA_t)+\sum_{ij}C_2\sqrt{-g}g^{tt}g^{ii}g^{jj}\partial_j\partial_j(\partial_tA_{ii}+\partial_i\partial_iA_t)\\
    0&=C_1\sqrt{-g}g^{tt}g^{ii}g^{jj}\partial_t(\partial_tA_{ij}+\partial_i\partial_jA_t)+C_1\partial_r(\sqrt{-g}g^{jj}g^{ii}g^{rr}\partial_rA_{ij}) \notag \\
    &+3C_3\sqrt{-g}g^{ii}g^{jj}\sum_k g^{kk} \partial_k(2\partial_kA_{ij}-\partial_iA_{jk}-\partial_jA_{ki})\notag\\&+ \delta_{ij}(C_2\partial_r(\sqrt{-g}g^{ii}g^{rr}\sum_kg^{kk}\partial_rA_{kk})+C_2\sqrt{-g}g^{tt}g^{ii}\sum_kg^{kk}\partial_t(\partial_tA_{kk}+\partial_k\partial_kA_t))
\end{align}
\end{subequations}
Let us now study the solutions to these equations of motion close to the boundary $r\rightarrow 0$, assuming a metric of the form \eqref{789} in the UV.  As $r\rightarrow 0$, terms with $\partial_r$ dominate over terms with $\partial_t$ or $\partial_i$, so we find that
\begin{subequations}\label{9}
\begin{align}
    A_{ij} &= A^{(0)}_{ij}+A^{(1)}_{ij}r^{-4+d+z-\theta+\frac{4\theta}{d}}+\ldots\\
    A_t &= A^{(0)}_t+A^{(1)}_tr^{d-z-\theta+\frac{2\theta}{d}}+\ldots
\end{align}
\end{subequations}
We will work under the assumptions that
\begin{subequations}\label{2001}
\begin{align}
    -4+d+z-\theta+\frac{4\theta}{d}>0\\
    d-z-\theta+\frac{2\theta}{d}>0
\end{align}
\end{subequations}
so that we can safely assume the terms $A^{(0)}_{ij}$ and $A^{(0)}_t$ to be our sources; and avoid complications. 

Other literature has studied models where the analogues of the criteria (\ref{2001}) are violated \cite{Klebanov_1999, Marolf_2006,Hartnoll:2009ns, Davison_2019}.  These authors argue that when one treats the non-constant term as source, the dual operator in the boundary theory no longer represents a conserved current, but rather a boundary gauge field.  Hence, we always wish to keep the constant $A^{(0)}$ terms as our sources, as we are interested in the dynamics of a conserved charge. For more  details on what kind of boundary terms to add if \eqref{2001} is violated, see the appendix.   

One possibility (which we will not consider in detail) is that the IR scaling theory, with exponents violating (\ref{2001}), transitions at some finite energy scale to a UV theory respecting (\ref{2001}).\footnote{Note that this transition could be due to the presence of other fields, in which case the backreaction of $A$ could be neglected (and complications discussed in Section \ref{lkj}) ignored).}  In this case, one might expect that the leading order term in the IR is the source field.  We expect that matching methods similar to those used for $\mathrm{AdS}_2$-Reissner-Nordstrom geometries (see e.g. \cite{hartnoll2016holographic}) could be useful for this scenario.

For now assuming (\ref{2001}), the action becomes a boundary term when evaluated on a solution to \eqref{1001}. Following the holographic renormalization prescription \cite{de_Haro_2001}, we introduce a UV cut-off at the boundary $r=\epsilon$ to study the behavior of the action and ensure its regularity.  We find that 
\begin{equation}\label{17}
\begin{split}
  S_{\mathrm{reg}} = \frac{1}{R^{\frac{(d-4)\theta}{d}}}\int \mathrm{d}^{d+1}x \epsilon^{3-d-z+\theta-\frac{2\theta}{d}}\bigg(
  -\epsilon^{2z-2}R^{-\frac{2\theta}{d}}A_t'A_t+C_1\epsilon^{2-\frac{2\theta}{d}}\sum_{ij}A_{ij}'A_{ij}+C_2\epsilon^{2-\frac{2\theta}{d}}\sum_jA_{jj}'\sum_iA_{ii}\bigg)
\end{split}
\end{equation}
Happily, this action is well-behaved even as $\epsilon\xrightarrow[]{}0$, so we don't need to add any counter terms. According to the holographic renormalization prescription, we can extract the expectation value of the current density operators as the coefficients of $S$ proportional to the sources.  We find that
\begin{subequations}
\begin{align}
    \langle J^{ij} \rangle &= \lim_{\epsilon\xrightarrow[]{}0}\frac{\delta S}{\delta A^{(0)}_{ij}},\\
    \langle J^{t} \rangle &= \lim_{\epsilon\xrightarrow[]{}0}\frac{\delta S}{\delta A^{(0)}_t},
\end{align}
\end{subequations}
and hence
\begin{subequations}\label{18}
\begin{align}
    \langle J^{ij} \rangle &= \frac{1}{R^{\frac{(d-4)\theta}{d}}}\left(-4+d+z-\theta+\frac{4\theta}{d}\right)\left(C_1A^{(1)}_{ij}+C_2\delta_{ij}\sum_{k}A^{(1)}_{kk}\right),\\
    \langle J^{t} \rangle &= -\frac{1}{R^{\frac{(d-2)\theta}{d}}}\left(d-z-\theta+\frac{2\theta}{d}\right)A^{(1)}_{t}.
\end{align}
\end{subequations}
\section{Transport and subdiffusion}
\subsection{Conductivity}
First, we calculate the direct current (zero frequency) conductivity which can be obtained by applying a uniform electric field.  Recall the discussion of the nature of conductivities in a dipole conserving system, presented in Section \ref{sec:subdifffield}.
To evaluate $\sigma^{ij,kl}$ holographically, we look for solutions to the equations of motion of the form
\begin{equation}\label{21}
\begin{split}
    A_{ij} &= (-E_{ij}t+a_{ij}(r)).\\
\end{split}
\end{equation}
 Plugging  into \eqref{1001} gives 
\begin{equation}
\begin{split}
    \partial_r\left(\sqrt{-g}g^{rr}\left(g^{xx}\right)^2(C_1+C_2\delta_{ij})\partial_rA_{ij}\right) &= 0.
\end{split}
\end{equation}
Hence, the object inside the parentheses is independent of the value of $r$.  Combining with \eqref{18}, we observe that as $r\rightarrow 0$, the object is equal to $\langle J^{ij}\rangle$: \begin{equation}
    \langle J^{ij} \rangle = \sqrt{-g}g^{rr}(g^{xx})^2(C_1\delta_{il}\delta_{jm}+C_2\delta_{ij}\delta_{lm})\partial_rA_{lm},
\end{equation}
while as $r\rightarrow r_+$, 
\begin{equation}\label{157}
\begin{split}
    \langle J^{ij}\rangle = \left.\sqrt{-g}g^{rr}(g^{xx})^2(C_1\delta_{il}\delta_{jm}+C_2\delta_{ij}\delta_{lm})a_{lm}'\right|_{r=r_+}.\\
\end{split}
\end{equation}
A quick way to ensure regularity of the solution at the horizon is to (temporarily) switch to Eddington-Finkelstein coordinate $v = t + \int \frac{\sqrt{g_{rr}}}{\sqrt{-g_{tt}}}\mathrm{d}r$.  Regularity of the solution then means that at the horizon, $a_{ij}$ should be a function of $v$ alone.  In other words,
\begin{equation}\label{22}
\begin{split}
    \partial_rA_{ij}|_{r=r_+} &= -\frac{\sqrt{g_{rr}}}{\sqrt{-g_{tt}}}\partial_tA_{ij}|_{r=r_+}
\end{split}
\end{equation}
Thus using \eqref{22} and \eqref{157} we find
\begin{equation}\label{800}
\begin{split}
    \sigma^{ij,lm} = &=  (C_1\delta_{il}\delta_{jm}+C_2\delta_{ij}\delta_{lm})r^{4-d}_+\bigg(\frac{r_+}{R}\bigg)^{\frac{(d-4)\theta}{d}}
\end{split}
\end{equation}

Note that the dc conductivity is written purely in terms of the horizon data in (\ref{157}). This is a consequence of the membrane paradigm \cite{1986bhmp.book.....T, Iqbal_2009}, which states that the transport in the boundary theory can be thought of as equivalently taking place on a fluid flowing across the horizon. The membrane paradigm also doesn't prohibit the conductivity from being dependent on $\theta$. 

\subsection{Subdiffusion constants}
In this section we try to compute the sub-diffusion constant by looking for plane wave perturbations to the gauge fields of the form
\begin{subequations}
\begin{align}
    A_t = a_t(r)\mathrm{e}^{-\mathrm{i}\omega t+\mathrm{i}kx}\\
    A_{ij} = a_{ij}\mathrm{e}^{-\mathrm{i}\omega t+\mathrm{i}kx}
\end{align}    
\end{subequations}
where we  have assumed  without loss of generality the momentum to be in the x-direction. Working with the background metric tensor given by \eqref{789}, the equations given in \eqref{1001} reduce to 
\begin{subequations}\label{741}
\begin{align}
    0&=r^{2z+\frac{2\theta}{d}}\omega a_t'(r)+\mathrm{i}k^2r^4f(a_{xx}'(r)(C_1+C_2)+C_5a_{yy}'(r))\\
    0&=\frac{dk^2r^3(k^2a_t(r)(C_1+C_2)+\mathrm{i}\omega(a_{xx}(r)C_1+(a_{xx}(r)+a_{yy}(r))C_2))}{f} + r^{\frac{2\theta}{d}}((d^2+2\theta-d(1+z+\theta))a_t'(r)-dra_t''(r))\\
    0&=\frac{dr^{2z}\omega(\omega (a_{xx}(r)C_1+(a_{xx}(r)+a_{yy}(r))C_2)-\mathrm{i}k^2a_t(r)(C_1+C_2))}{f}+dr(r((C_1+C_2)a_{xx}'(r)+C_2a_{yy}'(r))f'\notag\\&+f(C_1((5+\theta)a_{xx}'(r)+ra_{xx}''(r))+C_2((5+\theta)(a_{xx}'(r)+a_{yy}'(r))+r(a_{xx}''(r)+a_{yy}''(r)))))\notag\\&-r(d(d+z)+4\theta)f((C_1+C_2)a_{xx}'(r)+C_2a_{yy}'(r))\\
    0&=\frac{r^{2z}\omega^2a_{yy}(r)C_1}{f}-6k^2r^2a_{yy}(r)C_3+\frac{(d-1)r^{2z}\omega(\omega(a_{xx}(r)+a_{yy}(r))-\mathrm{i}k^2a_t(r))C_2}{f}\notag\\&+\frac{rC_1(a_{yy}'(r)(-(d(-5+d+z-\theta)+4\theta)f+drf')+drfa_{yy}''(r))}{d}-\notag\\&\frac{d-1}{d}rC_2(f((d(-5+d+z-\theta)+4\theta)(a_{xx}'(r)+a_{yy}'(r))-dr(a_{xx}''(r)+a_{yy}''(r)))-dr(a_{xx}'(r)+a_{yy}'(r))f')
\end{align}
\end{subequations}
The other components decouple from the calculation of subdiffusion constants and we will not consider them further.
Since finding an exact solution to this set of equations is not feasible, we try to find solutions where $\omega\ll T$, which is our area of interest. In order to study this problem, following \cite{Grozdanov_2019,Lucas:2015vna,Chen:2017dsy}, we split the holographic direction into three regions, namely: inner, outer, and intermediate regions. The outer region includes the near boundary ($r\xrightarrow[]{}0$) regime, where the solutions appear static (independent of $\omega$ and $k$), since the holographic direction corresponds to the energy scale, and extends inwards to a distance of $r_+-T^{-1}\mathrm{e}^{-4\pi T/\omega}$ (where our perturbative solution with $\omega=k=0$ fails). The inner region is close to the horizon and extends into the bulk upto a distance of $r_+-r \lesssim T^{-1}$. Here we impose in-falling boundary conditions to find the solutions. We then match the solutions in the intermediate region which exists between $r_+-T^{-1}<r<r_+-T^{-1}\mathrm{e}^{-4\pi T/\omega}$. 
\subsubsection{Outer region}
In the outer region which is close to $r\xrightarrow{}0$, the effect of $\omega,k$ are  negligible as can be seen from \eqref{741}, since no component of $a$ diverges. Let 
\begin{subequations}\label{770}
\begin{align}
    a_t(r,x^\mu) &= a^{(0)}_t(x^\mu) + j_t(x^\mu)\Phi_t(r)+\mathcal{O}(\omega,k)\\
    a_{xx}(r,x^\mu) &= a^{(0)}_{xx}(x^\mu) + j_{xx}(x^\mu)\Phi_{xx}(r)+\mathcal{O}(\omega,k)\\
    a_{yy}(r,x^\mu) &= a^{(0)}_{yy}(x^\mu) + j_{yy}(x^\mu)\Phi_{yy}(r)+\mathcal{O}(\omega,k)
\end{align}
\end{subequations}
Plugging the above ansatz into \eqref{741} we find 
\begin{subequations}
\begin{align}
    \Phi_t(r) &= \frac{r^{d-z-\theta+\frac{2\theta}{d}}}{d-z-\theta+\frac{2\theta}{d}}\\
    \Phi_{xx}(r) &= \Phi_{yy}(r) =  \int^{r}_{0} \mathrm{d}s\frac{s^{-5+d+z-\theta+\frac{4\theta}{d}}}{f(s)}
\end{align}
\end{subequations}
Comparing with \eqref{18} and replacing $j_t$ and  $j_{ii}$ in terms of $\langle J^t\rangle$ and  $\langle J^{ii}\rangle$ gives us 
\begin{subequations}
\begin{align}
    j_t &= - \langle J^t\rangle R^{\frac{(d-2)\theta}{d}}\\
    j_{ii} &= R^{\frac{(d-4)\theta}{d}}\bigg(\frac{\langle J^{ii}\rangle}{C_1}-\frac{C_2\sum_k\langle J^{kk}\rangle}{(C_1+dC_2)C_1}\bigg)
\end{align}
\end{subequations}
We find that $\Phi_{ii}$, with $i$ being the spatial indices, has a logarithmic  divergence at $r\xrightarrow{}r_+$ since $f(r)\xrightarrow{}4\pi Tr^{z-1}_+(r_+-r)+\mathcal{O}(r_+-r)^2$. We assume $d+z-\theta+\frac{4\theta}{d}>4$ to prevent UV divergences.
So we can rewrite the above equations by absorbing the divergence in a separate term as follows:   
\begin{subequations}
\begin{align}
    \Phi_t(r) &= \phi_t(r)\\ 
    \Phi_{ii}(r) &= \phi_{ii}(r) + \frac{r^{-5+d+z-\theta+\frac{4\theta}{d}}_+}{f'(r_+)}\log f(r)
\end{align}
\end{subequations}
where $\phi_t$, $\phi_{ii}$ are the finite parts
\begin{equation}
    \phi_{ii}(r) = \int^r_{0}\mathrm{d}s\frac{s^{-5+d+z-\theta+\frac{4\theta}{d}}}{f(s)}\bigg(1-\frac{f'(s)r^{-5+d+z-\theta+\frac{4\theta}{d}}_+}{s^{-5+d+z-\theta+\frac{4\theta}{d}}f'(r_+)}\bigg)
\end{equation}
\subsubsection{Inner region}
This regime is close to $r\xrightarrow{}r_+$. To solve the equations in this regime, we lets assume the gauge fields to be of form
\begin{subequations}\label{750}
\begin{align}
    a_t(r,x^\mu) &= \mathcal{A}^{(1)}_t(r,x^\mu)+\mathcal{A}^{(2)}_t(r, x^\mu)f(r)^{-\mathrm{i}\omega/4\pi T}\\
    a_{ii}(r,x^\mu) &= \mathcal{A}^{(1)}_{ii}(r,x^\mu)+\mathcal{A}^{(2)}_{ii}(r, x^\mu)f(r)^{-\mathrm{i}\omega/4\pi T}
\end{align}
\end{subequations}
where we have imposed the infalling boundary conditions for the second term. Since we are dealing with objects which are not gauge  invariant, we have the first term, which depends on the gauge. Now to solve for $\mathcal{A}_t, \mathcal{A}_{ii}$, we plug \eqref{750} into \eqref{741} and set the coefficients of diverging terms (such as $f^{-1},f^{-1-\mathrm{i}\omega/4\pi T}$) to zero at the horizon. This constraints our near horizon solution, leading to 
\begin{subequations}\label{742}
\begin{align}
    0&=\partial_x\partial_x \mathcal{A}^{(1)}_t(r_+, x^\mu)+\partial_t \bigg(\mathcal{A}^{(1)}_{xx}(r_+,x^\mu)+\frac{C_2}{C_1+C_2}\mathcal{A}^{(1)}_{yy}(r_+,x^\mu)\bigg)\\
    0&=\partial_x\partial_x \mathcal{A}^{(1)}_t(r_+, x^\mu)+\partial_t \bigg(\mathcal{A}^{(1)}_{xx}(r_+,x^\mu)+\bigg(1+\frac{C_1}{(d-1)C_2}\bigg)\mathcal{A}^{(1)}_{yy}(r_+,x^\mu)\bigg)\\
    0&=\mathcal{A}^{(2)}_t(r_+,x^\mu)\\
    0&=\partial_x\partial_x\partial_t\bigg(\mathcal{A}^{(2)}_{xx}(r_+,x^\mu)+\frac{C_2}{C_1+C_2}\mathcal{A}^{(2)}_{yy}(r_+,x^\mu)\bigg)
\end{align}
\end{subequations}
The first two equations can be used to fix $\mathcal{A}^{(1)}_{yy}(r_+,x^\mu)=0$. 
\subsubsection{Intermediate region}
In this region, the solutions corresponding to inner and outer region are both valid; hence we can match them. Since $\omega \ll T$, we can approximate
\begin{equation}
    f(r)^{-\mathrm{i}\omega/4\pi T}  \approx 1+\frac{\log f(r)}{4\pi T}\partial_t+\mathcal{O}(\partial^2)
\end{equation}
This makes \eqref{750} close to the horizon
\begin{subequations}\label{760}
\begin{align}
    a_t(r_+,x^\mu) &= \mathcal{A}^{(1)}_t(r_+,x^\mu)\\ 
    a_{xx}(r_+,x^\mu) &= \mathcal{A}^{(1)}_{xx}(r_+,x^\mu) + \mathcal{A}^{(2)}_{xx}(r_+,x^\mu)+\frac{\partial_t\mathcal{A}^{(2)}_{xx}(r_+,x^\mu)}{4\pi T}\log f(r)\\
    a_{xy}(r_+,x^\mu) &= \mathcal{A}^{(2)}_{xy}(r_+,x^\mu)+\frac{\partial_t\mathcal{A}^{(2)}_{xy}(r_+,x^\mu)}{4\pi T}\log f(r)\\
    a_{yy}(r_+,x^\mu) &= \mathcal{A}^{(2)}_{yy}(r_+,x^\mu)+\frac{\partial_t\mathcal{A}^{(2)}_{yy}(r_+,x^\mu)}{4\pi T}\log f(r)
\end{align}
\end{subequations}
Now  matching the finite and diverging parts of \eqref{760} and \eqref{770} gives
\begin{subequations}\label{743}
\begin{align}
    \partial_t\mathcal{A}^{(2)}_{ii}(r_+,x^\mu) &= \frac{4\pi Tr^{-5+d+z-\theta+\frac{4\theta}{d}}_+}{f'(r_+)}j_{ii}(x^\mu) = -r^{-4+d-\theta+\frac{4\theta}{d}}_+j_{ii}(x^\mu)\\
    \mathcal{A}^{(1)}_t(r_+,x^\mu) &= a^{(0)}_t(x^\mu)+j_t(x^\mu)\phi_t(r_+)\\
    \mathcal{A}^{(1)}_{xx}(r_+, x^\mu)+\mathcal{A}^{(2)}_{xx}(r_+,x^\mu) &= a^{(0)}_{xx}(x^\mu)+j_{xx}(x^\mu)\phi_{xx}(r_+)\\
    \mathcal{A}^{(2)}_{yy}(r_+,x^\mu) &= a^{(0)}_{yy}(x^\mu)+j_{yy}(x^\mu)\phi_{yy}(r_+)
\end{align}
\end{subequations}
Now we plug this into \eqref{742} to get the complete  set of  hydrodynamic equations to be 
\begin{subequations}\label{744}
\begin{align}
    0&= \partial_t\langle J^t\rangle + \partial_x\partial_x\langle J^{xx} \rangle\\
    -\frac{\phi_t(r_+)R^{\frac{2\theta}{d}}}{\phi_{xx}(r_+)}\partial_x\partial_x\langle J^t\rangle &= -\frac{R^{-\frac{(d-4)\theta}{d}}}{\phi_{xx}(r_+)}[\mathrm{d}a^{(0)}]_{x,t} - \frac{r^{-4+d-\theta+\frac{4\theta}{d}}_+}{\phi_{xx}(r_+)}\bigg(\frac{\langle J^{xx}\rangle}{C_1}-\frac{C_2\sum_k\langle J^{kk}\rangle}{(C_1+dC_2)C_1}\bigg) \label{744b}\\
    0 &= -\frac{R^{-\frac{(d-4)\theta}{d}}}{\phi_{yy}(r_+)}[\mathrm{d}a^{(0)}]_{yy,t}-\frac{r^{-4+d-\theta+\frac{4\theta}{d}}}{\phi_{yy}(r_+)}\bigg(\frac{(C_1+C_2)\sum_k\langle J^{kk}\rangle}{C_1(C_1+dC_2)}-\frac{\langle J^{xx}\rangle}{C_1}\bigg)
\end{align}
\end{subequations}
where \begin{equation}
    [\mathrm{d}a^{(0)}]_{ii,t}= \partial_i\partial_ia^{(0)}_t+\partial_ta^{(0)}_{ii}.
\end{equation}
Also, note that we have neglected time derivatives in the last two terms of (\ref{744}) since such terms are negligible in the $\omega\ll T$ limit.
Thus we obtain the subdiffusion constant
\begin{equation}\label{61}
    D = (C_1+C_2)\frac{r^{4-z-\frac{2\theta}{d}}_+R^{\frac{2\theta}{d}}}{(d-z-\theta+\frac{2\theta}{d})}.
\end{equation}
To find the dc susceptibility, we look at the solution for $a_t$ when $\omega=k=0$. This gives
\begin{equation} \label{eq:at}
    a_t = a^{(0)}_t(x^\mu)\bigg(1 - \bigg(\frac{r}{r_+}\bigg)^{d-z-\theta+\frac{2\theta}{d}}\bigg)
\end{equation}
where we have imposed the in-falling boundary condition $a_t(r_+) = 0$. But we know
\begin{equation}
    \chi = \partial_\mu\rho(\mu,T)
\end{equation}
where $\mu$ denotes the chemical potential which in our case is $a^{(0)}_t$. Combining with \eqref{18} we get
\begin{equation}
    \chi = \frac{(d-z-\theta+\frac{2\theta}{d})}{r^{d-z-\theta+\frac{2\theta}{d}}_+R^{\theta-\frac{2\theta}{d}}}
\end{equation} 
Einstein's relation implies that the conductivity
\begin{equation}
    \sigma_{xx,xx} = (C_1+C_2)r^{4-d}_+\bigg(\frac{r_+}{R}\bigg)^{\theta-\frac{4\theta}{d}}
\end{equation}
and we observe that this agrees with \eqref{800}. 
\subsection{Relation to butterfly velocity?}
The butterfly velocity quantifies the exponential growth of out-of-time-ordered correlation functions at long time and length scales \cite{Roberts:2014isa}. It was proposed in \cite{Blake_2016} that the butterfly velocity was a characteristic velocity for both relativistic and non-relativistic strongly interacting systems and that a diffusion bound (on conventional charge/energy diffusion) conjectured in \cite{Hartnoll_2014} could be expressed as
\begin{equation}
    D_2 \gtrsim \frac{v^2_{\mathrm{B}}}{T}
\end{equation}
where $D_2$ is the conventional second order charge diffusion constant, $\tau \sim \frac{\hbar}{k_{\mathrm{B}}T}$ is known as the `Planckian' time scale \cite{hartnoll2016holographic} and the model-dependent butterfly velocity $v_{\mathrm{B}}$  can be extracted purely from the horizon data \cite{Roberts:2014isa}.  Note that we are setting $\hbar=k_{\mathrm{B}}=1$.

Is it possible that in our model, the fourth order subdiffusion constant is also universal: \begin{equation}
    D_4 \sim \frac{v_{\mathrm{B}}^4 }{T^3}? \label{eq:D4bound}
\end{equation}
Unfortunately, the answer is no.   To compute $v_\mathrm{B}$ for our metric given in \eqref{789}, we follow the prescription of \cite{Blake_2016,Roberts_2016}, and find 
\begin{equation}
    v_\mathrm{B} = \frac{2\pi}{\beta} \left[ \frac{d\pi T\partial_r g_{xx}(r_+)}{r^{z-1}_+}\lim_{r\xrightarrow[]{} r_+}\frac{f(r)}{g_{tt}(r)}\right]^{-1/2}.
\end{equation}
For the hyperscaling violating metric (\ref{789}), we find \begin{equation}
  v_\mathrm{B} =    \frac{1}{r^{z-1}_+}\frac{\sqrt{d+z-\theta}}{\sqrt{2(d-\theta)}} \sim T^{1-\frac{1}{z}}.
\end{equation}

The sub-diffusion constant we have in \eqref{61} doesn't depend universally on $v_{\mathrm{B}}$ when $\theta\ne 0$, since the temperature dependence of $v_{\mathrm{B}}$ does not depend on $\theta$, while that of $D$ does. This does not imply the break down of the membrane paradigm: the sub-diffusion constant is still purely dependent on the horizon data.  What has happened is that some of the dimensionality of $D_4$ is made up of the constant $R$, rather than $T$, and so simple power counting alone does not fix the $T$ dependence of $D_4$ and $v_{\mathrm{B}}$ when $\theta \ne 0$.  (Observe that the horizon metric does depend on the scale $R$, and this is why $D_4$ has $R$ dependence.)  It is also possible for our inequalities in \eqref{2001} be violated for specific values of $d$, $z$ and $\theta$. In particular, if $a_t$ is largest near $r=0$ in (\ref{eq:at}), then the UV physics dominates the susceptibility $\chi$, so (\ref{eq:D4bound}) cannot hold since the right hand side depends only on near-horizon physics.  It may be the case that for another bulk action which (in the IR) describes the same subdiffusive physics, there is a more universal relation between subdiffusion constants and Planckian transport. 


\subsection{Finite frequency response}
Finally, we study the spatially homogeneous solutions to the bulk equations of motion at all frequencies.
\subsubsection{Conductivity at $\omega\xrightarrow{}\infty$ or $T=0$ limit}
We first analytically study the conductivity as $\omega\rightarrow\infty$.  As we will see, this limit also corresponds to $T\rightarrow 0$.  The equation of motion can be obtained from \eqref{1001}, and gives us (at $T=0$)
\begin{align}\label{988}
    (r^{5-d-z+\theta-\frac{4\theta}{d}}fa_{ij}')' &= -\frac{\omega^2a_{ij}}{fr^{d-3+\frac{4\theta}{d}-z-\theta}}
\end{align}
We remind the reader that equation (\ref{988}) was derived under the the condition that $C_4 = C_2$ and $C_5 = C_1$; for this reason, this equation of motion does not depend on any of the $C$ coefficients introduced earlier.
The solution to this equation which describes infalling modes is 
\begin{equation}
    a_{ij}(r) = c_{ij} r^{\frac{1}{2}(-4+d+z-\theta+\frac{4\theta}{d})}\mathrm{K}_{(-4+d+z-\theta+\frac{4\theta}{d})/2z}\bigg(-\frac{\mathrm{i}r^z\omega}{z}\bigg)
\end{equation}
with K the modified Bessel function.  The overall normalization constant is not important. If, say, we just source $a_{xx}$, then in the field theory this implies an electric field with non-vanishing $xx$-component.  To find the conductivity $\sigma_{xx,xx}$ we then Taylor series expand $a_{xx}$ about $r=0$ as  
\begin{equation}
    \lim_{r\xrightarrow{}0} a_{xx}(r) = r^{\frac{\alpha}{2}}\bigg(\mathrm{\Gamma}(\nu)\bigg(\frac{\mathrm{i}r^z\omega}{2z}\bigg)^{-\nu}+\mathrm{\Gamma}(-\nu)\bigg(\frac{\mathrm{i}r^z\omega}{2z}\bigg)^{\nu}+...\bigg)
\end{equation}
where \begin{subequations}
\begin{align}
    \alpha &= -4+d+z-\theta+\frac{4\theta}{d}, \\
    \nu &= \frac{\alpha}{2z}.
\end{align}
\end{subequations}
We also know that, for example,
\begin{equation}\label{111}
    \sigma_{xx,xx}(\omega) = \frac{G^{\mathrm{R}}_{J_{xx}J_{xx}}(\omega)}{\mathrm{i}\omega} = \frac{1}{\mathrm{i}\omega}\frac{\langle J^{xx}\rangle}{a^{(0)}_{xx}}
\end{equation}
We conclude that
\begin{subequations}\label{222}
\begin{align}
    \sigma_{xx,xx}|_{T=0} &= \frac{C_1+C_2}{\omega} \alpha(\frac{\omega}{2z})^{\frac{\alpha}{z}}\frac{\mathrm{\Gamma}(-\frac{\alpha}{2z})}{\mathrm{\Gamma}(\frac{\alpha}{2z})}\sin\left(\frac{\pi\alpha}{2z}\right)\\
    \sigma_{yy,xx}|_{T=0} &= \frac{C_2}{\omega} \alpha(\frac{\omega}{2z})^{\frac{\alpha}{z}}\frac{\mathrm{\Gamma}(-\frac{\alpha}{2z})}{\mathrm{\Gamma}(\frac{\alpha}{2z})}\sin\left(\frac{\pi\alpha}{2z}\right)
    \\    \sigma_{xy,xy}|_{T=0} &= \frac{C_1}{\omega} \alpha(\frac{\omega}{2z})^{\frac{\alpha}{z}}\frac{\mathrm{\Gamma}(-\frac{\alpha}{2z})}{\mathrm{\Gamma}(\frac{\alpha}{2z})}\sin\left(\frac{\pi\alpha}{2z}\right)
\end{align}
\end{subequations}
At finite $T$, so long as $\omega \gg T$, these response functions are still good approximations.

When $\theta=0$, we can understand the power law dependence in $\sigma(\omega)$ by general principles of dimensional analysis.  Starting from the assumption that $[\rho] = d$, $[t]=-z$ $[x]=-1$, using (\ref{eq:ward}) we find \begin{equation}
    [J_{ij}] = d+z-2.
\end{equation}
Using the formal definition of $\sigma_{ij,kl}$ in terms of Green's functions, we find that \begin{equation}
    [\sigma] = 2[J_{ij}] -(d+z)-z,
\end{equation}
where the first factor of $-(d+z)$ comes from the Fourier transform and the second $-z$ comes from the $\omega^{-1}$ factor in (\ref{111}).  We conclude that \begin{equation}
    \sigma(\omega) \sim \omega^{(d-4)/z}
\end{equation}
when $\theta=0$.   Our holographic calculation finds that when $\theta\ne0$, \begin{equation}
    \sigma(\omega)\sim \omega^{(d-4)(d-\theta)/dz}.
\end{equation}
Like in other holographic models \cite{hartnoll2016holographic}, it appears that our multipole-conserving holographic model, when $\theta\ne 0$, could only be consistent with a scaling theory where the charge density $\rho$ obtains an anomalous dimension \cite{Hartnoll:2015sea,Gouteraux:2013oca,Gouteraux:2014hca}, in conflict with the standard lore \cite{Sachdev_1994}, which has only been violated thus far in microscopically finely tuned models \cite{Karch:2015pha}.
\subsubsection{Conductivity at finite frequency and finite temperature}
Now, we numerically solve the equation \eqref{988} as a function of the frequency $\omega$.
 Once we do so, we can obtain the conductivity using \eqref{111}.  In our numerics, we have set $r_+=R=1$ for simplicity, and so $f(r) = 1 - r^{d+z-\theta}$.

\begin{figure}[t]
\includegraphics[width=0.49\textwidth]{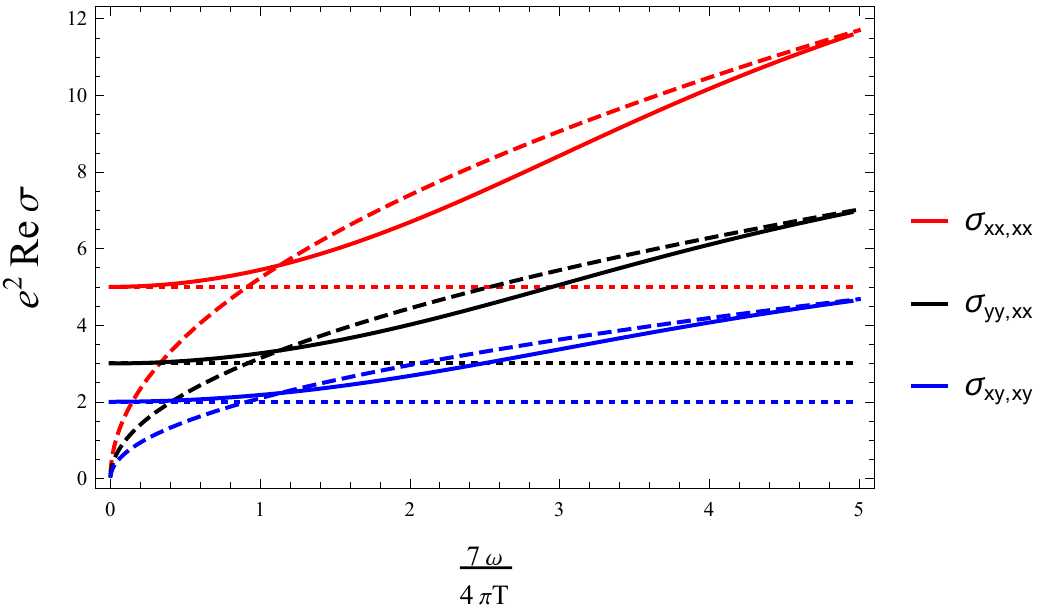}
\includegraphics[width=0.49\textwidth]{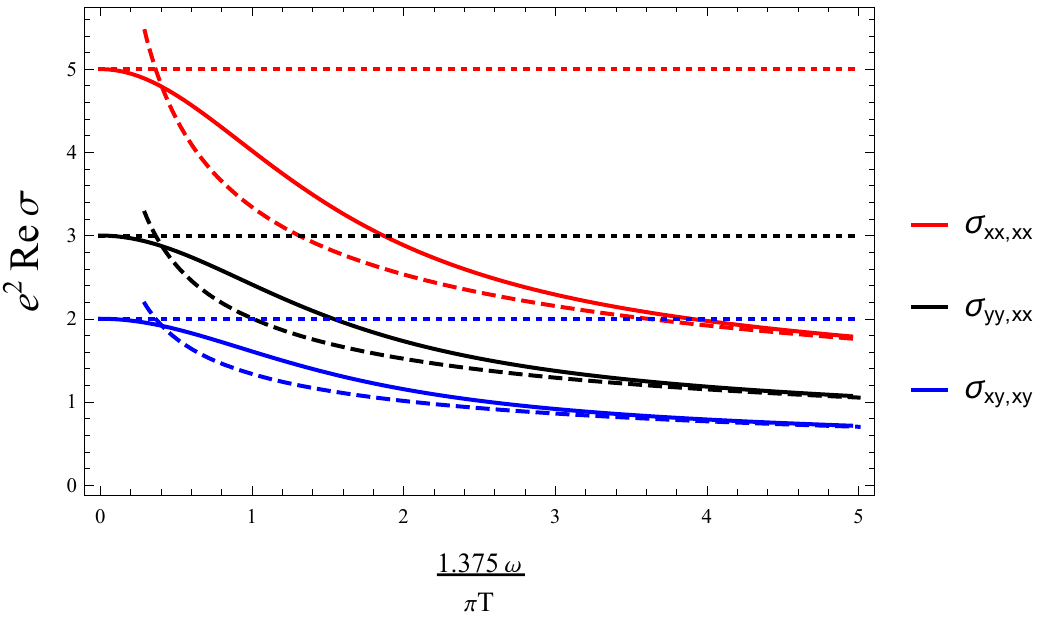}
  \caption{Real part of frequency-dependent conductivity for the theory described by \eqref{3} (dotted and dashed lines corresponds to the limiting cases $\omega=0$ and $\omega\xrightarrow[]{}\infty$ respectively). Left: Conductivity contribution from each channel for $d=5$, $\theta=0$, $z=2$, $C_1=2$, $C_2=3$. Right:  Conductivity contribution from each channel  for $d=3$, $\theta=0$, $z=\frac{5}{2}$, $C_1=2$, $C_2=3$.}
  \label{fig}
\end{figure}
\FloatBarrier
In Figure \ref{fig}, we see that this conductivity matches the dc $(\omega=0$) value which we analytically obtained. It also matches with our analytic result in \eqref{222} at $\omega \gg T$. 

\subsubsection{Conductivity independent of frequency}
Lastly, we show that when $d=4$, the conductivity is independent of frequency, even at finite temperature $T$.
With a little bit of hindsight \cite{hartnoll2016holographic}, we see that if we want a frequency independent conductivity, we require equation \eqref{988} to yield solutions of the form
\begin{equation}
    a_{ii}(r) = \exp\bigg(\mathrm{i}\omega\int^r_{0}\mathrm{d}s \frac{s^\alpha}{f(s)}\bigg) \label{eq:specialsoln}
\end{equation}
which gives a frequency independent $\sigma$.
Plugging (\ref{eq:specialsoln}) into (\ref{988}) imposes the condition
\begin{equation}
    5-d-z+\theta-\frac{4\theta}{d} = d-z-3-\theta+\frac{4\theta}{d}
\end{equation}
The physically acceptable solution to this equation is $d=4$.

In the case of an ordinary charge conserving theory, without dipole conservation, in $d=2$ the simplest holographic models predict a frequency-independent conductivity.  This effect is reminiscent of  particle-vortex duality \cite{Herzog:2007ij}.

\section{Conclusion}

In this paper, we have introduced a simple holographic model that allows for the study of multipole-conserving dynamics.  Our results are in agreement with the recently proposed theory of hydrodynamics in such systems \cite{gromov2020fracton}.

In the future, we hope our framework can be generalized in a number of directions.  First and foremost, it would be interesting to understand whether there is an alternative covariant bulk construction of a boundary-multipole-conserving theory which does not break translation invariance.  The solution to this problem appears related to the challenge of coupling mixed-rank tensors to gravity, and may have broader implications for the formal study of fracton matter.  Secondly, it would be interesting to understand more complex features of the ($\omega,k$)-dependent conductivity \cite{Pinchpoint}, which may be one of the more practical ways to look for (approximate) fractonic matter in experiments.  Finally, we do not know whether or not our model can be related in any way to other recent attempts \cite{Yan:2018nco} to link fracton matter with holography, and it would be interesting to understand this point further.

\section*{Acknowledgements}
We  thank Paolo Glorioso and Michael Pretko for useful discussions.  AL is supported by a Research Fellowship from the Alfred P. Sloan Foundation.

\appendix
\section{Holographic renormalization}
If \eqref{2001} is violated, we need to add boundary counterterms to the bulk action to make it finite.  As in the main text, we proceed by imposing a hard cutoff on the bulk at radius $r=\epsilon$, and demand that the action is finite as $\epsilon \rightarrow 0$. We focus on theories with $\theta=0$, where the gauge-invariant boundary action takes the form
\begin{equation}\label{178}
    S_{\mathrm{bdy}} = -\frac{1}{2}\int \mathrm{d}^{d+1}x \sqrt{-\gamma}\bigg(\frac{g^{rr}g^{tt}(F^1)^2}{d-z}+\frac{C_1g^{rr}g^{ii}g^{jj}(F^3_{ij})^2+C_2g^{rr}g^{ii}F^3_{ii}g^{jj}F^3_{jj}}{-4+d+z}\bigg)
\end{equation}
where $\gamma$ is the induced metric at $r=\epsilon$.
The renormalized action would then be given by 
\begin{equation}
    S_{\mathrm{renorm}} = S_{\mathrm{reg}}+S_{\mathrm{bdy}}
\end{equation}
with $S_{\mathrm{reg}}$ the action given in (\ref{3}). Varying both the boundary term and the regulated action gives us
\begin{subequations}
\begin{align}
  \delta S_{\mathrm{reg}} &= \int \mathrm{d}^{d+1}x \bigg(
  -(d-z)A^{(1)}_t(\delta A^{(0)}_t+\delta A^{(1)}_t\epsilon^{d-z})+(-4+d+z)(C_1\sum_{ij}A^{(1)}_{ij}(\delta A^{(0)}_{ij}+A^{(1)}_{ij}\epsilon^{-4+d+z})\nonumber\\&\hspace{25em}+C_2\sum_jA^{(1)}_{jj}\sum_i(\delta A^{0}_{ii}+A^{(1)}_{ii}\epsilon^{-4+d+z}))\bigg)\\
  \delta S_{\mathrm{bdy}} &= \int \mathrm{d}^{d+1}x \bigg(
  (d-z)A^{(1)}_t\delta A^{(1)}_t\epsilon^{d-z}-(-4+d+z)(C_1\sum_{ij}A^{(1)}_{ij}A^{(1)}_{ij}\epsilon^{-4+d+z}+C_2\sum_jA^{(1)}_{jj}\sum_iA^{(1)}_{ii}\epsilon^{-4+d+z}))\bigg)
\end{align}
\end{subequations}
where the higher order terms go to zero as $\epsilon\xrightarrow[]{}0$.
This renormalized action now becomes finite as $\epsilon\xrightarrow[]{}0$: 
\begin{align}
    \delta S_{\mathrm{renorm}} &= \delta S_{\mathrm{reg}} + \delta S_{\mathrm{bdy}} \notag \\
    &= \int \mathrm{d}^{d+1}x \left(-(d-z)A^{(1)}_t\delta A^{(0)}_t+(-4+d+z)\left(C_1\sum_{ij}A^{(1)}_{ij}\delta A^{(0)}_{ij}+C_2 \sum_{i}A^{(1)}_{ii}\sum_{j}\delta A^{(0)}_{jj}\right)\right)
\end{align}
Now we can use this to compute 1-point functions of the dual operator
\begin{subequations}
\begin{align}
    \langle J^t\rangle = \frac{\delta S_{\mathrm{renorm}}}{\delta A^{(0)}_t}\\
    \langle J^{ij} \rangle = \frac{\delta S_{\mathrm{renorm}}}{\delta A^{(0)}_{ij}}
\end{align}
\end{subequations}
This gives us the formulas equivalent to \eqref{18}.

We can also add an alternate boundary term
\begin{align}\label{132}
    S_{\mathrm{bdy}} = \frac{1}{2}\int \mathrm{d}^{d+1}x \sqrt{-\gamma}\bigg(\frac{g^{rr}g^{tt}(F^1)^2}{d-z}+\frac{C_1g^{rr}g^{ii}g^{jj}(F^3_{ij})^2+C_2g^{rr}g^{ii}F^3_{ii}g^{jj}F^3_{jj}}{-4+d+z}\bigg)\nonumber\\ - \int \mathrm{d}^{d+1}x \sqrt{-\gamma}n_r(g^{rr}g^{tt}F^1A_t+C_1g^{rr}g^{ii}g^{jj}F^3_{ij}A_{ij}+C_2g^{rr}g^{ii}F^3_{ii}g^{jj}F^3_{jj})
\end{align}
where we have flipped the sign of the first term and added a second term. Here $n$ is an outward pointing unit normal to the surface $r=\epsilon$. This also leads to a finite total action whose variation now takes the form
\begin{subequations}
\begin{align}
  \delta S_{\mathrm{reg}} &= \int \mathrm{d}^{d+1}x \bigg(
  -(d-z)A^{(1)}_t(\delta A^{(0)}_t+\delta A^{(1)}_t\epsilon^{d-z})+(-4+d+z)(C_1\sum_{ij}A^{(1)}_{ij}(\delta A^{(0)}_{ij}+A^{(1)}_{ij}\epsilon^{-4+d+z})\nonumber\\&\hspace{25em}+C_2\sum_jA^{(1)}_{jj}\sum_i(\delta A^{0}_{ii}+A^{(1)}_{ii}\epsilon^{-4+d+z}))\bigg)\\
  \delta S_{\mathrm{bdy}} &= \int \mathrm{d}^{d+1}x \bigg(
  (d-z)(A^{(1)}_t\delta A^{(0)}_t+A^{(0)}_t\delta A^{(1)}_t+ A^{(1)}_t\delta A^{(1)}_t\epsilon^{d-z})-(-4+d+z)(C_1\sum_{ij}(A^{(0)}_{ij}\delta A^{(1)}_{ij}\nonumber\\&+A^{(1)}_{ij}\delta A^{(0)}_{ij}+A^{(1)}_{ij}A^{(1)}_{ij}\epsilon^{-4+d+z})+C_2\sum_j\sum_i(A^{(0)}_{ii}\delta A^{(1)}_{jj}+A^{(1)}_{jj}\delta A^{(0)}_{ii}+A^{(1)}_{jj}A^{(1)}_{ii}\epsilon^{-4+d+z})))\bigg)
\end{align}
\end{subequations}
Thus,
\begin{align}
    \delta S_{\mathrm{renorm}} &= \delta S_{\mathrm{reg}} + \delta S_{\mathrm{bdy}} \notag \\
    &= \int \mathrm{d}^{d+1}x \left((d-z)A^{(0)}_t\delta A^{(1)}_t-(-4+d+z)\left(C_1\sum_{ij}A^{(0)}_{ij}\delta A^{(1)}_{ij}+C_2 \sum_{i}A^{(0)}_{ii}\sum_{j}\delta A^{(1)}_{jj}\right)\right)
\end{align}
Therefore the roles of $A^{(0)}$ and $A^{(1)}$ are exchanged. This gives us 
\begin{subequations}
\begin{align}
    \langle J^t_{\mathrm{alt}}\rangle &= \frac{\delta S_{\mathrm{renorm}}}{\delta A^{(1)}_t} = \left(d-z\right)A^{(0)}_{t}\\
    \langle J^{ij}_{\mathrm{alt}} \rangle &= \frac{\delta S_{\mathrm{renorm}}}{\delta A^{(1)}_{ij}} = -\left(-4+d+z\right)\left(C_1A^{(0)}_{ij}+C_2\delta_{ij}\sum_{k}A^{(0)}_{kk}\right)
\end{align}
\end{subequations}
The dimensions of the dual operators are $[J^t_{\mathrm{alt}}] = z$ and $[J^{ij}_{\mathrm{alt}}] = 2$. But this is dimensionally inconsistent with \eqref{eq:ward} which can be seen as a consequence of the fact that the second term in \eqref{132} breaks gauge invariance. It has been argued in \cite{Marolf_2006, Argurio_2017} that for the regular $\mathrm{U}(1)$ bulk gauge field the dual operator is another gauge field as  opposed to the usual current-density. We expect a similar situation for  our case.  

For our theory, when $\theta \neq 0$, we find the following counterterm:
\begin{align}
    S_{\mathrm{bdy}} = -\frac{1}{2}\int \mathrm{d}^{d+1}x \sqrt{-\gamma}n_r(g^{rr}g^{tt}F^1A_t+C_1g^{rr}g^{ii}g^{jj}F^3_{ij}A_{ij}+C_2g^{rr}g^{ii}F^3_{ii}g^{jj}F^3_{jj})
\end{align}
Since 
\begin{subequations}
\begin{align}
  \delta S_{\mathrm{reg}} &= \frac{1}{R^{\frac{(d-4)\theta}{d}}}\int \mathrm{d}^{d+1}x \bigg(
  -R^{-\frac{2\theta}{d}}\bigg(d-z-\theta+\frac{2\theta}{d}\bigg)A^{(1)}_t(\delta A^{(0)}_t+\delta A^{(1)}_t\epsilon^{d-z-\theta+\frac{2\theta}{d}})\nonumber\\&+\bigg(-4+d+z-\theta+\frac{4\theta}{d}\bigg)(C_1\sum_{ij}A^{(1)}_{ij}(\delta A^{(0)}_{ij}+\delta A^{(1)}_{ij}\epsilon^{-4+d+z-\theta+\frac{4\theta}{d}})+C_2\sum_j\sum_iA^{(1)}_{jj}(\delta A^{(0)}_{ii}+\delta A^{(1)}_{ii}\epsilon^{-4+d+z-\theta+\frac{4\theta}{d}})\bigg)\\
  \delta S_{\mathrm{bdy}} &= \frac{1}{2R^{\frac{(d-4)\theta}{d}}}\int \mathrm{d}^{d+1}x \bigg(
  R^{-\frac{2\theta}{d}}\bigg(d-z-\theta+\frac{2\theta}{d}\bigg)(A^{(1)}_t\delta A^{(0)}_t+A^{(0)}_t\delta A^{(1)}_t+ 2A^{(1)}_t\delta A^{(1)}_t\epsilon^{d-z-\theta+\frac{2\theta}{d}})\nonumber\\&\hspace{10em}-\bigg(-4+d+z-\theta+\frac{4\theta}{d}\bigg)(C_1\sum_{ij}(A^{(0)}_{ij}\delta A^{(1)}_{ij}+A^{(1)}_{ij}\delta A^{(0)}_{ij}+2A^{(1)}_{ij}A^{(1)}_{ij}\epsilon^{-4+d+z-\theta+\frac{4\theta}{d}})\nonumber\\&\hspace{15em}+C_2\sum_j\sum_i(A^{(0)}_{ii}\delta A^{(1)}_{jj}+A^{(1)}_{jj}\delta A^{(0)}_{ii}+2A^{(1)}_{jj}A^{(1)}_{ii}\epsilon^{-4+d+z-\theta+\frac{4\theta}{d}})))\bigg),
\end{align}
\end{subequations}
the renormalized action becomes
\begin{align}
    \delta S_{\mathrm{renorm}} &= \delta S_{\mathrm{reg}} + \delta S_{\mathrm{bdy}} \notag \\
    &= \frac{1}{2R^{\frac{(d-4)\theta}{d}}}\int \mathrm{d}^{d+1}x \bigg(
  -R^{-\frac{2\theta}{d}}\bigg(d-z-\theta+\frac{2\theta}{d}\bigg)(A^{(1)}_t\delta A^{(0)}_t-A^{(0)}_t\delta A^{(1)}_{t})\nonumber\\&+\bigg(-4+d+z-\theta+\frac{4\theta}{d}\bigg)(C_1\sum_{ij}(A^{(1)}_{ij}\delta A^{(0)}_{ij}-A^{(0)}_{ij}\delta A^{(1)}_{ij})+C_2\sum_j\sum_i(A^{(1)}_{jj}\delta A^{(0)}_{ii}-A^{(0)}_{ii}\delta A^{(1)}_{jj})\bigg)
\end{align}
It appears to us that the possible boundary counterterms which render a finite action when (\ref{2001}) is violated, and $\theta=0$, break gauge invariance, since a term analogous to \eqref{178} has a variation
\begin{align} \label{eq:a13}
  \delta S_{\mathrm{reg}} &= -\frac{\epsilon^{-\frac{\theta}{d}}}{R^{-\frac{\theta}{d}}R^{\frac{(d-4)\theta}{d}}}\int \mathrm{d}^{d+1}x \bigg(
  -R^{-\frac{2\theta}{d}}\bigg(d-z-\theta+\frac{2\theta}{d}\bigg)A^{(1)}_t\delta A^{(1)}_t\epsilon^{d-z-\theta+\frac{2\theta}{d}}\nonumber\\&+\bigg(-4+d+z-\theta+\frac{4\theta}{d}\bigg)(C_1\sum_{ij}A^{(1)}_{ij}\delta A^{(1)}_{ij}\epsilon^{-4+d+z-\theta+\frac{4\theta}{d}}+C_2\sum_j\sum_iA^{(1)}_{jj}\delta A^{(1)}_{ii}\epsilon^{-4+d+z-\theta+\frac{4\theta}{d}}\bigg)
\end{align}
which does not cancel the divergence in $\delta S_{\mathrm{reg}}$ due to the overall prefactor of $\epsilon^{-\theta/d}$ in (\ref{eq:a13}).

\bibliography{thebib.bib}
\end{document}